\documentclass[aps,floatfix,preprint,nofootinbib]{revtex4}

\usepackage{graphicx}
\usepackage{epic}
\usepackage{eepic}
\usepackage{latexsym}

\newcommand{\eq}[1]{(\ref{#1})}
\newcommand{\be}{\begin{equation}}
\newcommand{\ee}{\end{equation}}
\newcommand{\bea}{\begin{eqnarray}}
\newcommand{\eea}{\end{eqnarray}}

\newcommand{\vs}[1]{\vspace{#1 mm}}
\newcommand{\hs}[1]{\hspace{#1 mm}}

\def\a{\alpha}
\def\b{\beta}

\def\d{\delta}

\def\e{\epsilon}

\def\f{\phi}
\def\fr{\frac}

\def\k{\kappa}
\def\l{\lambda}

\def\m{\mu}
\def\n{\nu}

\def\r{\rho}
\def\s{\sigma}

\def\O{\Omega}

\def\o{\omega}

\def\del{\partial}

\let\bm=\bibitem
\def\nn{\nonumber}
\newcommand{\cn}{{\cal N}}

\begin{document}

\title{Open Strings on D-Branes and\\ Hagedorn Regime in String Gas Cosmology}

\author{Ay\c{s}e Arslanarg\i n$^{1}$}
\email[]{ayse.aslanargin@boun.edu.tr}
\author{Ali Kaya$^{1,2}$\vs{3}}
\email[]{ali.kaya@boun.edu.tr}
\affiliation{$^{1}$Bo\~{g}azi\c{c}i University, Department of Physics, \\ 34342,
Bebek, \.Istanbul, Turkey\vs{3}\\
$^{2}$Feza G\"{u}rsey Institute,\\
Emek Mah. No:68, \c{C}engelk\"{o}y, \.Istanbul, Turkey\vs{3}}

\date{\today}

\begin{abstract}

We consider early time cosmic evolution in string gas cosmology dominated by open strings attached to D-branes. After reviewing statistical properties  of open strings in D-brane backgrounds, we use dilaton-gravity equations to determine the string frame fields. Although, there are distinctions in the Hagedorn regime thermodynamics and  dilaton coupling as compared to closed strings, it seems difficult to avoid Jeans instability and assume thermal equilibrium simultaneously, which is already a known problem for closed strings.  We also examine characteristics of a possible subsequent large radius regime in this setup. 

\end{abstract}

\maketitle

\section{Introduction}

As a candidate for a unified theory of all interactions and quantum gravity, string theory still lacks any direct or indirect observational support. One hopes that cosmology may alter this situation and offer a testing ground for string theory or any other quantum theory of gravity in not so distant future. Therefore, it is important to study possible cosmological implications of string theory. String gas cosmology \cite{bv} is one possible approach which takes into account not only the massless low energy fields but all stringy excitations like winding modes, which are expected to play important roles both in early and late time cosmologies (see \cite{rev1,rev2} for recent reviews). 

Except in special situations usually preserving some supersymmetry, we do not have an understanding of string theory in the strong coupling regime. For cosmologically relevant time dependent backgrounds this lack of knowledge prevents proper understanding of the key issues like the resolution of big-bang singularity.  However, this does not stop one to study  the theory in the weak coupling limit. Indeed,  string theory has a very rich structure even when the interactions are weak and this motivates the study of  toy cosmological models. For instance, it is known that  for strings at very high energies there exists a "limiting" Hagedorn temperature  and one wonders possible implications of this feature in cosmology. On the other hand, studies in the string/brane gas cosmology reveal that stringy excitations offer mechanisms for late time stabilization of extra dimensions (see e.g. \cite{ex1,ex2,ex3,ex4,ex5,ex6,ex7,ex8,ex9}), shape moduli (see e.g. \cite{sh1,sh2,sh3,sh4}) and dilaton (see e.g.  \cite{ex9,dil1,dil2, dil3}).

Since the thermal partition function does not converge for temperatures above the Hagedorn temperature, thermodynamics of strings should be studied in the microcanonical ensemble. There are subtleties in taking the thermodynamical limit in this formalism \cite{tan1,tan2,tan3}, but assuming a totally compact space, the microcanonical approach offers a suitable framework in studying string cosmology near "big-bang". To be specific, one can use basic thermodynamical relations to deduce the right hand side of the Einstein's equations and study the cosmological evolution. In string theory, at least in the weak coupling limit, all the necessary thermodynamical variables are already determined  and ready to use for cosmology. For closed strings, this approach was used in \cite{s1,s2,s3,s4,s5,s6,s7,s8} to study different aspects of string gas cosmology. In \cite{s7}, however, an important obstacle regarding the assumption of thermal equilibrium in early time string gas cosmology is pointed out. Namely, it is noticed that the interaction rates for closed strings turn out to be small compared to expansion rate and thus strings cannot keep thermal equilibrium, which is an important barrier in such scenarios. 

Our aim in this paper is to study an early time cosmology dominated by open strings attached to D-branes. In the absence of any physical, mathematical or philosophical constraints about initial conditions in the big-bang, it is natural to expect the existence of D-branes even at early times. Of course, according to Brandenberger-Vafa (BV) mechanism \cite{bv} higher dimensional D-branes are expected to annihilate, but such arguments rely on the assumption of thermal equilibrium which are subject to discussion as noted above. Indeed, for D-branes the interaction rates are studied in \cite{dbrane}, which indicates difficulties for BV mechanism applied to D-branes. 

In the presence of D-branes, it is known that open strings dominate the system in thermal equilibrium. Physically,  this can be understood as a result of D-branes chopping up the closed strings \cite{b1,b2,abel}. In the micro-canonical approach, the statistical properties of open strings on D-brane backgrounds is determined in \cite{abel} and there are key differences compared to closed strings. Besides, the coupling of open strings to dilaton alters gravity field equations in a non-trivial way. Thus, one expects a different cosmological evolution in the presence of D-branes. Our aim is to check whether these differences may help to allow the assumption of thermal equilibrium for open strings in a simple setting. It turns out that this problem also persists for open strings in the weak coupling limit and thus studies of string gas cosmology with D-branes also require understanding of non-equilibrium thermodynamics or strong coupling dynamics or both.  

The plan of the paper is as follows. In the next section, following \cite{abel}, we review thermodynamics of open strings on parallel D-branes which are uniformly distributed along some compact directions. In section \ref{sec3}, we use this thermodynamical information in dilaton-gravity field equations to determine the early time cosmological evolution. We check the consistency of the background and examine the reliability  of the thermal equilibrium. In this section, we also study characteristics of a  possible subsequent large radius regime. We conclude with brief remarks and possible future directions in section \ref{sec4}.

\section{Review: The Entropy of open strings in D-brane backgrounds} \label{sec2}

\begin{figure}
\centerline{
\includegraphics[width=7.5cm]{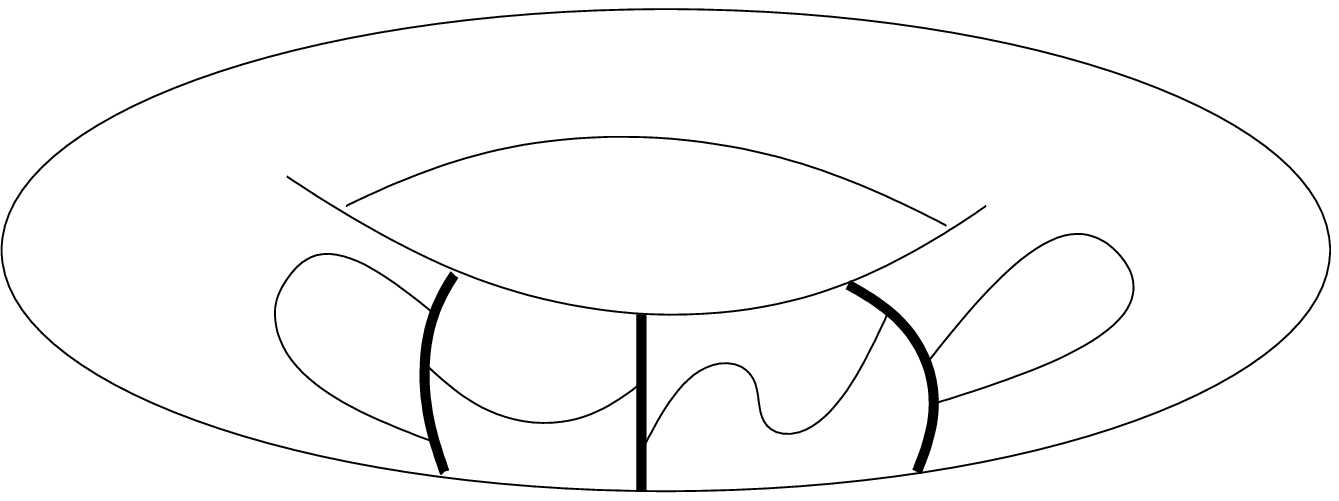}}
\caption{Parallel D-branes in a compact space and open strings attached to them.}.
\label{fig1}
\end{figure}

In this section, we review the calculation of the entropy of open strings attached to D-branes in a totally compact toroidal space at temperatures close to the Hagedorn temperature and in the weak string coupling limit. The main results derived from first principles can be found in \cite{abel} and here we simply repeat the derivations in a slightly different way for the cases we are interested in (see also \cite{b1,b2} for an approach based on Boltzmann equations). We assume that there are $n$ parallel $d_N$ dimensional D-branes in the system which are distributed uniformly in  $d_D$ dimensional transverse directions, where $d_N+d_D=9$ (see figure \ref{fig1}). Actually, in the absence of any symmetry one expects to have zero net RR charge and thus an equal number of D-branes and anti-D-branes in the system. In the next section, we consider this possibility, but for now we simply assume only the existence of D-branes. We use $R_N$ and $R_D$ to denote the corresponding radii for the world-volume and transverse directions. The end points of open strings are imposed to obey Neumann boundary conditions along the world-volume directions and Dirichlet  boundary conditions along the transverse directions, which are labeled by $x$ and $y$ coordinates, respectively. The space-time metric can be written as
\be
ds^2=-dt^2+R_N^2(d\vec{x})^2+R_D^2(d\vec{y})^2,
\ee
and it will be convenient to define the volumes of the Neumann and Dirichlet directions 
\be
V_N=(R_N)^{d_N}, \hs{5} V_D=(R_D)^{d_D}.
\ee
We take all dimensions to be larger than string scale 
\be
R_N\geq 1,\hs{5} R_D \geq  1,
\ee
since one can apply a T-duality transformation to a small direction to make it large (we set $\a'=1$).

An open string in such a background is labeled by $d_N$ momentum numbers, $d_D$ winding numbers, the oscillator level and the two Chan-Paton factors identifying the D-branes on which the end points of the string are attached. Note that, unlike closed strings, there are no momentum modes along Dirichlet and no winding modes along Neumann directions. In the micro-canonical ensemble and for $\e\gg1$, the single string density of states is given by (see e.g. \cite{abel})
\be\label{w}
\o(\e)=f\, e^{\b_h \e}\, \sum_{l}\, g_l\, e^{-l^2 \e/\e_0},
\ee
where the sum is over the integers $0\leq l<R_D$, $\b_h=2\sqrt{2}\pi$ is the Hagedorn temperature for open strings, $\e_0=R_D^2$, $g_l$ is a degeneracy factor\footnote{Indeed, $g_l$  counts the number of integer $d_D$-tuples whose length is equal to $l$, see \cite{abel}.} which is approximately given by $g_l\simeq 2 \textrm{Vol}(S^{d_D-1})l^{d_D-1}$ for $l\gg 1$ and  
\be\label{f}
f=n^2\fr{V_N}{V_D}.
\ee
Recall that $n$ is the number of D-branes in the system and $n^2$ factor in \eq{w} arises from the combinatorics of the Chan-Paton factors.

The form of the single string density of states depends on the magnitude of the string energy $\e$ relative to the moduli parameter $\e_0$ \cite{abel}:
\be\label{wcases}
\o(\e)\simeq \begin{cases}{  f\, e^{\b_h \e},\hs{23} \e\gg\e_0,\cr
f\, \left(\e_0/\e\right)^{d_D/2}\,e^{\b_h \e},\hs{3}\e\ll\e_0.}\end{cases}
\ee
For $\e\gg \e_0$ only the first term contributes in the sum in \eq{w} and in this limit the strings are energetic enough to wind around the Dirichlet directions. For example, this is the regime of interest for $R_D={\cal O}(1)$ and $\e\gg1$. In the opposite limit $\e\ll \e_0$, the sum can be replaced by an integral and in this case the available energy is not large enough to excite the winding modes. 

The total density of states $\O(E)$, which counts the degeneracy with fixed energy $E$ and unconstrained number of particles, is given by 
\be
\O(E)=\sum_k\frac{1}{k!}\,\prod_{i=1}^{k}\,\int_0^E\, \o(\e_i)\,d\e_i\,\d\left(\sum_i\e_i-E\right).
\ee
Using the integral representation of the delta function, $\O(E)$ can be expressed as
\be\label{W}
\O(E)=\fr{1}{2\pi E}\,\int_{-\infty}^{\infty}\,d\a\, e^{-i\a}e^{F(\a)},
\ee
where 
\be\label{df}
F(\a)=\int_0^{E}\,d\e\,\o(\e)\,e^{i\a\,\e/E}.
\ee
To get the leading order asymptotic contribution one can use \eq{w} in \eq{df} to obtain
\be\label{f2}
F(\a)\simeq fE\sum_l\,g_l\,\fr{e^{i\a+E\b_l}-1}{i\a+E\b_l},
\ee
where the temperature $\b_l$ is  defined as 
\be
\b_l=\b_h-\fr{l^2}{\e_0}.
\ee
Note that $F(\a)$ is a regular function which vanishes as $\a\to\pm\infty$ and therefore the integral \eq{W} is convergent. 

\begin{figure}
\centerline{
\includegraphics[width=7.5cm]{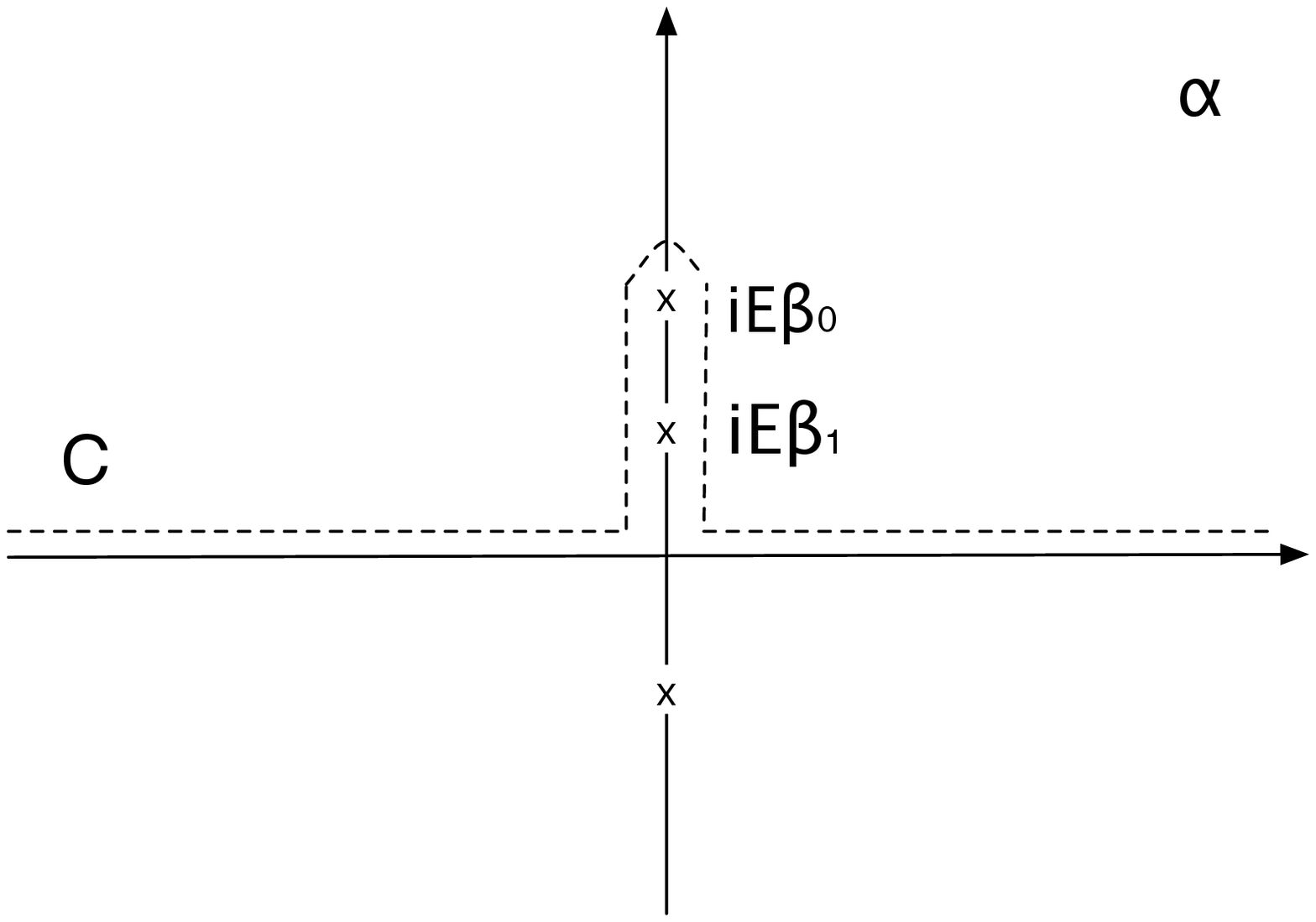}}
\caption{The deformed integration contour $C$, which is used to calculate the integral \eq{W}.}
\label{fig2}
\end{figure}

To calculate \eq{W}, we use a method which is  similar to the complex temperature formalism \cite{tan1,tan2,tan3}. Since $F(\a)$ is an analytic function in the complex $\a$-plane (note that the points on the imaginary axis $\a=iE\b_l$  are regular) one can deform the integral contour near the origin by extending it  through the imaginary axis to circle around the point $iE\b_0$, as shown in  figure \ref{fig2}. The analytic function $F(\a)$ can be written as the sum of two non-analytic functions $F(\a)=F_1(\a)+F_2(\a)$, where $F_1$ and $F_2$ are the pieces containing the exponential and $(-1)$ in the denominator of \eq{f2}, respectively. By expanding $\textrm{exp}(F_1)$ in powers of $F_1$ one can see that only the first term  in the expansion contributes to the integral along the deformed contour. This is because, in evaluating all but the first term the contour can be closed from above by excluding the poles of $F_1$ and $F_2$  and the integral along  the upper semi-circle at infinity vanishes. Therefore we have 
\be\label{c}
\O(E)\simeq\fr{1}{2\pi E}\,\int_C\,d\a\, \textrm{exp}\left(-i\a-fE\sum_l \fr{g_l}{i\a+E\b_l}\right),
\ee
where the contour $C$ is pictured in figure \ref{fig2}. This line integral can now be calculated by closing the contour from below; since there is no contribution coming from the lower semi-circle the integral equals to the sum of the residues at $\a=iE\b_l$ which we denote by $\O_l(E)$:
\be\label{cs}
\O(E)=\sum_l\,\O_l(E).
\ee
It is easy to see that $\O_l(E)\sim\exp(\b_l E)$, therefore \eq{cs} offers an expansion scheme which can be seen to be equivalent to the complex temperature formalism \cite{tan1,tan2,tan3}. In the following we concentrate on  the calculation of $\O_0(E)$, the first term in this expansion . Since the number of poles depends on the  radius of the Dirichlet directions, one should consider two different regimes depending on the magnitude of $R_D$. 

\subsection{Small radius regime}  

In the small radius regime, i.e. when $R_D={\cal O}(1)$, it is enough to consider only  $l=0$ term in the sum in \eq{c}, and thus we have\footnote{As a convention, we define $\oint$ to denote the line integral over a small circle around the singularity of the integrand, which is suitable for the calculation of the residue.}
\be\label{c2}
\O(E)\simeq\O_0(E)=\fr{1}{2\pi E}\,\oint\,d\a\,\textrm{exp}\left(-\fr{fE}{i\a+E\b_h}\right)\, e^{-i\a}.
\ee
Expanding the first exponential and evaluating the residue at $\a=iE\b_h$ term by term one finds \cite{abel}
\bea
\O(E)&\simeq&e^{\b_hE}\sum_k\fr{(fE)^k}{k!(k-1)!}\fr{1}{E}=\sqrt{\fr{f}{E}}\,\textrm{exp}\left(\b_hE\right)\,I_1\left(2\sqrt{fE}\right)\nn\\
&\simeq&\fr{f^{1/4}}{E^{3/4}}\,\textrm{exp}\left(\b_h E+2\sqrt{fE}\right) ,\hs{6}E\gg1,\label{o1}
\eea
where $I_1$ is the modified Bessel function of first kind. Compared to the closed string density of states, the main difference is the  presence of the squareroot term in the exponential in \eq{o1}. Note that \eq{o1} is valid even when $R_N\gg1$. Indeed for larger $R_N$ the asymptotic expansion becomes more accurate.

Although the residue can exactly be calculated in this case, one can also apply a saddle point approximation to extract the asymptotic behavior. After defining $\a=iE\b_h+z \,\sqrt{fE}$ the integral \eq{c2} becomes 
\be
\O(E)\sim e^{\b_h E} \sqrt{\fr{f}{E}}\oint dz\, e^{-i\sqrt{fE}\,(z-1/z)}.
\ee
For $fE\gg1$, which is automatically satisfied in the small radius regime since $V_D={\cal O}(1)$, one can use a saddle point approximation about the point $z=i$ which precisely gives \eq{o1}.

\subsection{Large radius regime}

If $R_D\gg1$, the $l\not=0$ terms should also be taken into account in the sum in \eq{c}  while evaluating $\O_0(E)$. This makes the exact calculation of the residue very difficult. However, under suitable conditions, a saddle point approximation can be used to get the asymptotic behavior. In terms of a new complex variable $z=-(i\a+\b_hE)/E$, the density function $\O_0(E)$ becomes
\be\label{s1}
\O_0(E)=\fr{e^{\b_h E}}{2\pi i}\oint dz\,\exp\left(Ez+f/z+\s(z)\right),
\ee
where $\s(z)$ is defined by
\be
\s(z)=f\sum_{l>0}\fr{g_{l}}{z+l^2/\e_0}=\sum_{k\geq0}\,\s_k\,z^k.
\ee
Note that $\s(z)$ is an analytic function near $z=0$ so that it can be expanded in a Taylor series as above. Using the fact that  $g_l\simeq 2 \textrm{Vol}(S^{d_D-1})l^{d_D-1}$ for $l\gg 1$ the expansion coefficients can be found as
\be
\s_k\,=\,(-1)^k a_k\,f\,V_D, 
\ee
where $a_k$ is a positive constant of order unity. With the help of this expansion and further defining 
\be
\k=\sqrt{f/(E+\s_1)},\hs{5}\,\l=\sqrt{f(E+\s_1)},\hs{5}z=\k\, y,
\ee
\eq{s1} can be converted into 
\be\label{s}
\O_0(E)\simeq e^{\b_hE+\s_0}\,\k\,\oint dy \exp\left(\l\left[y+1/y+\sum_{k\geq2}\tilde{\s}_ky^k\right]\right),
\ee
where $\tilde{\s}_k=\k^k\,\s_k/\l$. Assuming that $\l\gg1$, the saddle point approximation is applicable to evaluate \eq{s}. Moreover, if  $\tilde{\s}_k\ll1$ for $k\geq2$, $y=-i$ becomes the saddle point up to small controllable  corrections. Under these assumptions the saddle point calculation gives \cite{abel}
\be\label{lo}
\O_0(E)\simeq \fr{f^{1/4}}{(E-a_1 f V_D)^{3/4}}\,\exp\left(\b_h E+a_0\,f\,V_D+2\sqrt{f(E-a_1\,f\,V_D)}\right),
\ee
which is valid when $\l\ll1$ and $\tilde{\s}_k\ll1$ for $k\geq2$. One can see that these two conditions are equivalent to
\be\label{con}
E\gg f V_D+\fr{1}{f}.
\ee 
For $f\geq 1$, \eq{con} is identical to $E\gg f V_D$. 

The contribution of the $l$'th pole $\O_l(E)$ can be carried out in a similar fashion ($z$-variable should now be defined as $-E\,z=i\a+\b_l\,E$). One can see that if the saddle point approximation is applicable to $\O_0(E)$, i.e. when \eq{con} holds, it is also pertinent in the evaluation of $\O_l(E)$. A straightforward calculation then gives
\be
\O_l(E)\simeq \fr{f^{1/4}}{\left[g_l(E-\tilde{a}_1 f V_D)\right]^{3/4}}\,\exp\left(\b_l E+\tilde{a}_0\,f\,V_D+2\sqrt{g_l\,f\,(E-\tilde{a}_1\,f\,V_D)}\right),
\ee
where $\tilde{a}_0$ and $\tilde{a}_1$ are two other constants of order unity. If one demands that $\O_0\gg\O_1$, an additional condition on $E$ should be imposed 
\be\label{con2}
E\gg f\,R_D^2\,V_D\hs{3}\textrm{or}\hs{3}fR_D^4,
\ee
which is stronger than \eq{con} for $f\geq1$. For $d_D\geq2$, the first term becomes larger than the second term in the right hand side of the above inequality. As discussed in \cite{abel}, if \eq{con2} is not satisfied one has to change the form of the single string density of states $\o(\e)$ in the calculation of $\O(E)$ according to \eq{wcases}. 

\section{The dilaton-gravity background}\label{sec3}

In this section we use dilaton-gravity equations sourced by open strings attached to D-branes to determine the string frame fields. Before deriving the field equations, let us discuss why such a setup might be relevant in the context of string gas cosmology. The original scenario proposed in \cite{bv} has been developed in \cite{bgas} to include all higher dimensional excitations in string theory. As it is natural in such settings, all degrees of freedom are assumed in thermal equilibrium in a hot and dense state. Then, by applying the BV mechanism to D$p$-branes of different dimensions, it is argued in \cite{bgas} that all higher dimensional branes  with $p>3$ annihilate in 10-dimensions and the remaining branes form a hierarchy of sizes of compact dimensions. Namely, 2-branes only permit a 5-dimensional subspace to grow, in which only 3-dimensions are allowed to expand by strings. 

As discussed in \cite{bgas}, even though BV mechanism works perfectly for higher dimensional branes, causality requires at least one brane per Hubble volume remaining. Although, a subsequent loitering phases can lead to a total annihilation \cite{loiter}, it is not very unnatural to assume that some D-branes survive even BV mechanism functions well. Besides, using Boltzmann equations the annihilation of D-branes in an expanding universe has been studied in \cite{dbrane}, which indicates that the BV mechanism may not work as efficient as one may think and higher dimensional D-branes may also fall out of thermal equilibrium surviving the annihilation. Therefore, it is worth to consider models with some D-branes left over in the system. 

At this point, one should notice that there is no conflict in assuming out of equilibrium D-branes and open strings in thermal equilibrium. In saying D-branes fall out of equilibrium, one refers to D-brane anti-D-brane annihilation process, which happens only when branes physically intersect each other. For instance, if there exists only a {\it single} D-brane in the universe as it is usually assumed in brane-world models, then BV mechanism is inapplicable, yet one can still think about open strings attached to that D-brane being in thermal equilibrium as it is studied in \cite{b2} or \cite{abel}.\footnote{It is interesting to note that when all directions are compact and at string size, the assumption of homogeneity can be justified even in the presence of a single D-brane, since in thermal equilibrium open strings exist in a  long string phase and they traverse the entire space many number of times before ending on the D-brane, which ensures homogeneity.}

As pointed out in the previous section if the net RR-charge of the universe is zero, then there must exist an equal number of D-branes and anti-D-branes. One then wonders whether the results of the previous section alter in such a modification. In \cite{b2}, the thermodynamics of the system has been studied in this more general setup using the Boltzmann equations and (under suitable conditions) the number of states accessible  with energy $E$ is found to be the same (see eq. (19) in \cite{b2}). On the other hand, in a realistic situation one expects to encounter generic intersections rather than just parallel D-branes. Although it is possible to motivate the sole existence of parallel D-branes in the context of brane-world models, we show at the end of this section that the main conclusion of this paper does not change if one considers generic intersections.

Let us now start discussing the dilaton gravity equations. At weak string coupling $g_s=e^\phi\ll1$, the action for the dilaton and the metric can be written as
\be\label{action}
S=\int d^{10}x\,\sqrt{-g}\,\,\left[\,\,
e^{-2\phi}(R\,+\,4\nabla\phi^2)\,+\,e^{-\f}\,
{\cal L}_o\right],\label{str} 
\ee
where ${\cal L}_o$ is the effective Lagrangian for open strings and the dilaton dependence is explicitly singled out in \eq{action}. The field equations following from this action can be found as  
\bea
&&R_{\m\n}+2\nabla_\m\nabla_\n\f-\fr{1}{2}\left[R+4\nabla^2\f-4
  (\nabla\f)^2\right]g_{\m\n}=e^{\f}\,T_{\m\n},\label{ff1}\\ 
&&R+4\nabla^2\f-4 (\nabla\f)^2+\fr{1}{2}e^\f\,{\cal L}_o=0,\label{ff2}
\eea
where $T_{\m\n}$ is the energy momentum tensor
\be
T_{\m\n}=-\fr{1}{\sqrt{-g}}\,\fr{\del}{\del
  g^{\m\n}} \left(\sqrt{-g}\,{\cal L}_o\right).
\ee
To proceed further one needs information about ${\cal L}_o$. It is noticed in \cite{ex4}  that imposing the conservation formula $\nabla_\m T^{\m\n}=0$ as the matter field equations, the contracted Bianchi identity can fix ${\cal L}_o$ in terms of the energy-momentum content, at least in a cosmological setting. Taking 
\be
\f=\f(t),\hs{4}g_{ti}=0,\hs{4}T_{ti}=0,
\ee
where $i$ labels a spatial direction, the consistency of \eq{ff1} and \eq{ff2} implies 
\be\label{los}
{\cal L}_m\,=\,-\r\, ,
\ee
which is precisely the Lagrangian for hydrodynamical matter. Therefore, given any conserved energy-momentum tensor, equations \eq{ff1}, \eq{ff2}  together with \eq{los}  determine the dynamics.

In our case, the supposed D-brane configuration suggests  
\be
ds^2=-dt^2+e^{2B(t)}(d\vec{x})^2+e^{2C(t)}(d\vec{y})^2,
\ee
where the radii of the Neumann and Dirichlet directions are given as
\be
R_N=e^{B},\hs{5}R_D=e^C.
\ee
Assuming $\f=\f(t)$ and $T_{\m\n}=\textrm{diag}(\r,p_N,p_D)$, the equations \eq{ff1} and \eq{ff2} can be shown to imply 
\bea
&&\ddot{B}+K\dot{B}=e^\f\left(p_N+\fr{\r}{2}\right),\nn\\
&&\ddot{C}+K\dot{C}=e^\f\left(p_D+\fr{\r}{2}\right),\nn\\
&&\ddot{\f}+K\dot{\f}=e^\f \left(\fr{d_N}{2}p_N+\fr{d_D}{2}p_D-\fr{3}{2}\r\right),\label{efe}\\
&&K^2=d_N\dot{B}^2+d_D\dot{C}^2+2\r\,e^{\f},\nn
\eea
where dot denotes derivative with respect to $t$ and  
\be
K\equiv d_N\dot{B}+d_D\dot{C}-2\dot{\f}.
\ee
Compared to evolution equations obtained for closed strings that have no dilaton coupling,  the system  \eq{efe} has a very crucial difference: there appears the energy density $\r$ in the right hand side of $\ddot{B}$ and $\ddot{C}$ equations. Therefore, the "force" along a direction, which determines its cosmic evolution, is not only given by the pressure as in the case of closed strings but it is equal to sum of the pressure with energy density. 

The energy-momentum tensor for open strings on D-brane backgrounds can be derived from the entropy of the system determined in the previous section. Given $S(E,V_N,V_D)$ one can find out  the temperature and the pressures as 
\be
\fr{1}{T}=\fr{\del S}{\del E},\hs{5}P_N=T\,V_N\,\fr{\del S}{\del V_N},\hs{5} P_D=T\,V_D\,\fr{\del S}{\del V_D}.
\ee
The densities that enter in the right hand side of \eq{efe} are given by
\be
\r=\fr{E}{V},\hs{5}p_N=\fr{P_N}{V},\hs{5}p_D=\fr{P_D}{V},
\ee
where $V=V_N\,V_D$. Assuming that the cosmic evolution is adiabatic, i.e.  $dS=0$, implies 
\be 
\dot{E}+d_N\dot{B}P_N+d_D\dot{C}P_D=0,
\ee
which is equivalent to conservation of energy-momentum tensor $\nabla_\m T^{\m\n}=0$.

In the small radius regime, from \eq{o1} one finds
\be\label{ss}
S\simeq \b_hE+2\sqrt{fE},
\ee
and 
\be\label{tp}
\fr{1}{T}=\b_h+\sqrt{\fr{f}{E}},\hs{5}P_N=T\sqrt{fE},\hs{5}P_D=-T\sqrt{fE},
\ee
which shows that the temperature is always smaller than the Hagedorn temperature and the pressures are very large. Not surprisingly, the pressures along Neumann and Dirichlet directions turn out to be positive and negative, respectively, which is due to the presence/absence  of momentum and winding modes. 

Assuming that the universe starts out at the string radii $B_0=C_0=0$ with an expansion $\dot{B}_0,\,\dot{C}_0>0$, one now has all the necessary  information to determine the early cosmic evolution in this toy model. In principle one should specify 6 initial conditions for $B$, $C$ and $\f$, constrained by the last equation in \eq{efe}. Also, the initial energy $E_0$ (or constant entropy $S$) and the number of D-branes $n$ in the system should be given. For $E_0\gg1$, equations \eq{ss} and \eq{tp} can be used to determine the right hand side of \eq{efe} until say $R_D\sim 3$. After $R_D>3$, one should consider the expressions in the large radius regime. 

We construct  a simple analytic but approximate solution valid in the small radius regime  as follows. Although the presence of D-branes breaks isotropy, we assume that initially $\dot{B}_0=\dot{C}_0$. We take $n={\cal O}(1)$ or so to avoid D-brane anti-D-brane annihilation (see the end of this section) and thus suppose 
\be
E_0\gg n^2.
\ee
Under these assumptions, one sees from \eq{tp} that $T\simeq 1/\b_h$ to a very good approximation. Moreover, the pressure terms in the right hand side of \eq{efe} can be neglected compared to the energy density. Therefore, one effectively gets a pressure-less phase. Indeed, solving $E$ from \eq{ss}
\be
\b_hE\simeq S+\fr{2f}{\b_h}-\fr{2}{\b_h}\sqrt{f(f+S\b_h)},
 \ee
one can see that it can approximately be treated as a constant (equal to $S/\b_h$) when $f$ is not changing appreciably since $S\gg f$. This is consistent with an effective pressure-less phase. 

Ignoring the pressures, one can set $B=C$ during the small radius regime. The system for $B$ and $\f$ can be rewritten in terms of the "conformal time" $\eta$ as
\bea
&&B''=\fr{1}{2}E\,e^{9B-3\f},\nn\\
&&\f''=\fr{3}{2}E\,e^{9B-3\f},\label{eq}\\
&&72B'^2+4\f'^2-36B'\f'=2Ee^{9B-2\f},\nn
\eea
where 
\be
e^{9B-2\f}d\eta=dt,
\ee
and prime denotes derivative with respect to $\eta$.

\begin{figure}
\centerline{
\includegraphics[width=5.2cm]{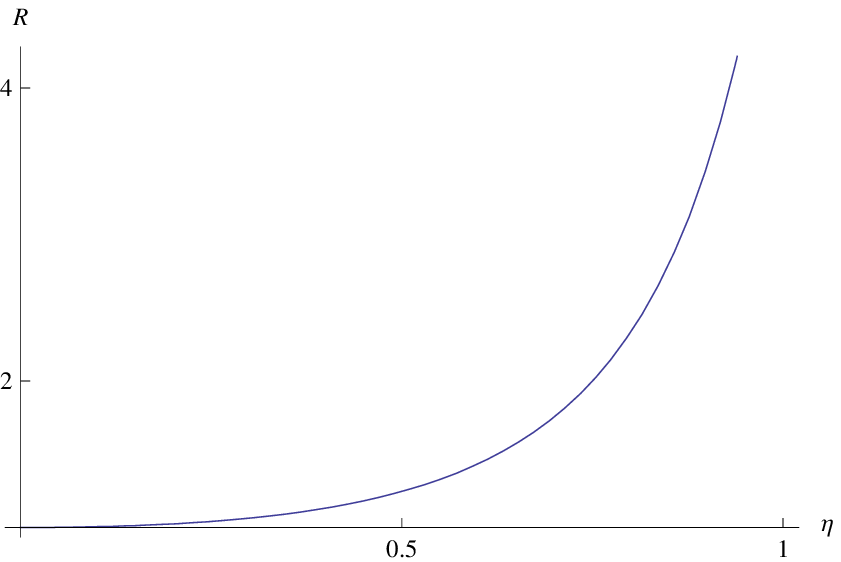}\hs{3}\includegraphics[width=5.2cm]{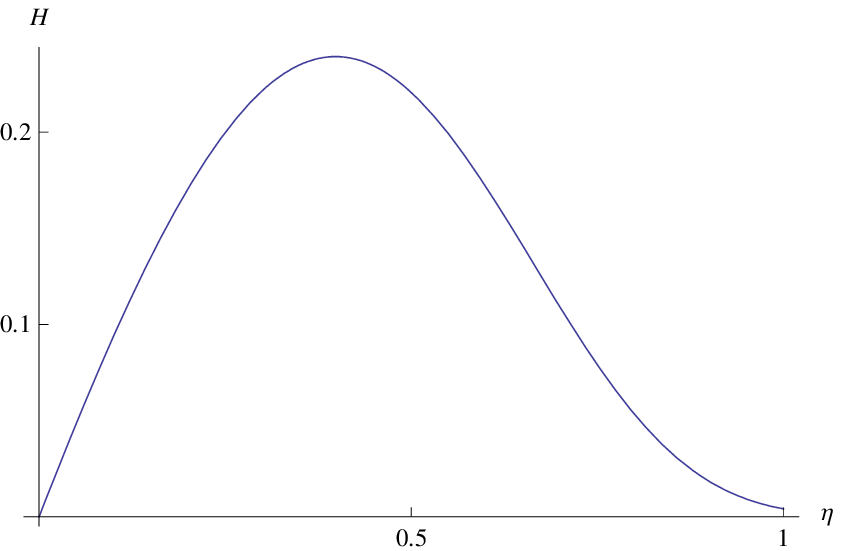}\hs{3}\includegraphics[width=5.2cm]{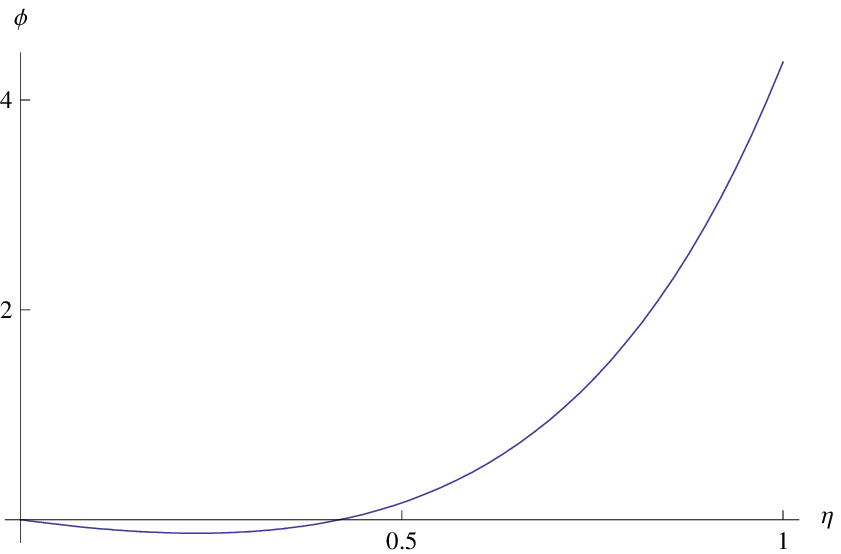}}
\caption{The plots of $R=e^{B}$, $H$ and $\f$ for $\l=1/9$. We take $c=1$ in the graphs.}
\label{fig3}
\end{figure}

The first two equations give $\f=3B+c\,\eta+\f_0$, which can be used to get a single second order differential equation for $B$. There appears two more constants of integration in the solution of $B$ one of which can be fixed in terms of the other constants using the constraint  equation in \eq{eq}.  Choosing $\eta=0$ to be the initial time and setting also $B(0)=0$ we find
\bea
&&B=\l\left(e^{3c\eta}-1\right)-\fr{c}{3}\,\eta,\label{sol1}\\
&&\f=3\l\left(e^{3c\eta}-1\right)-2c\,\eta+\f_0,\label{sol2}
\eea
where
\be
c^2=\fr{E}{18\l}\,e^{-3\f_0}
\ee
and $\l$ is a free positive constant. By picking  $\l$ to be positive, one makes sure that  the proper time $t$ increases with $\eta$. For $\l\geq1/9$, the dimensions continuously expand where the expansion speed initially vanishes for $\l=1/9$. For $\l<1/9$, the initial expansion speed is negative giving a period of contraction which later turns into an ongoing expansion. These can be seen from the Hubble expansion parameter  which can be determined as 
\be\label{hg}
H=\dot{B}=e^{-9B+2\f}B'=c\exp\left[-3\l(e^{3c\eta}-1)-c\,\eta+2\f_0\right]\left[3\l e^{3c\eta}-1/3\right].
\ee
Also note the initial Hubble rate 
\be\label{ih}
H_0=c\,e^{2\f_0}\left[3\l-1/3\right].
\ee
Recall that this solution can be trusted during the  small radius regime, i.e. until the time $\eta*$ such that $e^{B(\eta*)}\sim 3$. The change in dilaton during this period is also of order unity  $\f(\eta*)-\f(0) ={\cal O}(1)$. We give the plots of the functions $\exp(B)$, $\f$ and $H$ for $\l=1/9$ in figure \ref{fig3}.

Having obtained the solution, one can now check the consistency of the background. Firstly, we should make sure that the prescribed initial conditions avoid Jeans instability, i.e. the gravitational collapse. For a black hole not to form in a region of size $R$ in $d$-spatial dimensions, the mass inside this region $\r R^d$ should yield a  Schwarzschild radius\footnote{Note that the Schwarzschild radius corresponding to a mass $M$ in $d$-dimensions is given by $r_s=(GM)^{1/(d-2)}$.} smaller than $R$, which implies 
\be\label{j0}
R^2\leq \fr{1}{G\r},
\ee
where $G$ is the gravitational coupling constant. Although for closed strings $G=e^{2\f}$, the coupling of open strings to gravity is determined by  $G=e^{\f}$. In our case since $R=1$ at $t=0$, \eq{j0} gives
\be\label{co1}
e^{\f_0}\leq\fr{1}{E_0},
\ee
which constraints the initial value of  dilaton in terms of initial energy. Since the change in dilaton is of order unity,  Jeans instability will be avoided  at later times in the small radius regime, if it is avoided at $\eta=0$ with a margin. 

Secondly, we would like to  check whether the assumption of thermal equilibrium can be justified in this model. This is an important constraint which is known to be problematic for closed strings \cite{s7}. Thermal equilibrium requires the interaction rate per string $\Gamma$ to be larger than the expansion rate $H$:
\be\label{eqc}
\Gamma\geq H.
\ee
Although this condition  has already been used in the literature in dilaton-gravity setting (see, e.g. \cite{s7,s8}), let us try to explain why it is still valid, since it is known that some standard results valid in Einstein gravity are modified in dilaton gravity. Equation \eq{eqc} can be deduced from the Boltzmann equation for the number density $n$ of a species, which roughly takes the following form in $d$-dimensions: $\dot{n}+dHn+\Gamma n^2=0$ . The Hubble term arises since the number density decreases with the scale factor as $1/R^d$ and the last term comes from the integration of the interaction cross sections. It is clear that in dilaton gravity the form of this equation is the same, consequently \eq{eqc} can be used to justify thermal equilibrium. In a more fundamental setting, if one considers the Boltzmann equation for the number density in the phase space, then the Hubble term arises from the geodesic equation. In our case, the conservation of the energy momentum tensor $\nabla_\m T^{\m\n}=0$ ensures that strings move on the geodesics in the string frame, which again shows that \eq{eqc} can be used in dilaton gravity. 

We first evaluate $\Gamma$ without paying attention to a possible ``long'' string dominance. In that case, the rate can be estimated as
\be\label{r1}
\Gamma_{\textrm{short}}\sim \cn\,e^{2\f},
\ee
where $\cn$ is the number of strings per unit volume. To find $\cn$, one can define the average number $D(\e)$ of strings carrying a fixed amount of energy $\e$, which is given by \cite{tan2}
\be
D(\e)=\fr{\o(\e)\O(E-\e)}{\O(E)}.
\ee
Since $V={\cal O}(1)$, the total number of open strings also equals to the number density, which can be found as
\be\label{cnc}
\cn=\int_0^E\,D(\e)\,d\e\simeq\,\sqrt{fE},
\ee
where we used \eq{wcases} and \eq{o1}.  Therefore, \eq{r1} gives 
\be\label{sse}
\Gamma_{\textrm{short}}\sim n\,\sqrt{E}\,e^{2\f},
\ee
where presently  insignificant  $n$ dependence in the above formula is kept for future use. 

On the other hand, it is known that strings in a compact space near Hagedorn temperatures can exhibit a long string phase where a single string can traverse the entire space several times. In that case, the interaction rate \eq{r1} should be modified to take into account this fact. Assuming a  classical discrete model as discussed in \cite{hs1,hs2}, the interaction rate for a single string should be proportional to its length times the total length of the rest of the strings
\be\label{long}
\Gamma_{\textrm{long}}\sim e^{2\f}\bar{L}\left[(\cn-1)\bar{L}\right],
\ee 
where $\bar{L}$ is the average length of open strings. 

In our case, the average length $\bar{L}$ roughly equals to the average energy, which can then be found as 
\be\label{length}
\bar{L}\simeq\bar{\e}=\fr{E}{\cn}\simeq\sqrt{\fr{E}{f}}. 
\ee
For $n={\cal O}(1)$, the length becomes $\bar{L}\gg1$, which shows that the system is indeed dominated by long strings. Therefore the interaction rate can be estimated from \eq{long} as 
\be\label{irs}
\Gamma\sim e^{2\f}\,E^{3/2}.
\ee
Although the interaction rate is greatly enhanced compared to short string estimate \eq{sse}, equation \eq{co1} implies that $\Gamma\ll1$. 

Let us now check whether the assumption of thermal equilibrium can be justified. Using \eq{ih} and \eq{irs}, the condition \eq{eqc} at $\eta=0$  implies that  $\l$ should  be fine tuned in a small neighborhood of $1/9$. Recall that  when $\l=1/9$ the initial expansion speed vanishes, so this is not surprising. Therefore, with a fine tuning it is possible to justify the assumption of thermal equilibrium initially. However, from \eq{hg} the Hubble parameter can be seen to increase and reach a maximum value at around $c\eta\sim 0.4$. The Hubble parameter then decreases a little bit, but at the end of the short radius regime, which corresponds to $c\eta*\sim 1$, it becomes roughly equal to its maximum value which can be determined from \eq{hg} as
\be\label{hb}
H\sim \sqrt{E}\,e^{\f_0/2}.
\ee
Looking at the solution, one finds that the dilaton is not changing appreciably during this interval. Therefore \eq{eqc} implies 
\be\label{co31}
\fr{1}{E^{2/3}}\leq e^{\f_0}.
\ee
It is not possible to satisfy \eq{co31} along with \eq{co1} for $E\gg1$ and $e^{\f_0}\ll1$. Therefore, we see that even the initial conditions are fine tuned to avoid Jeans instability and justify  thermal equilibrium, the open string gas most likely fall out of thermal equilibrium during the small radius regime.

The only loophole in the above argument is that the estimated interaction rate $\Gamma$ can actually be larger due to some numerical factors of $\pi$'s or sums over spin or momentum states etc. If these factors become large enough then it might be possible to avoid Jeans instability and assume thermal equilibrium for certain energies smaller than a critical value. However, this looks both difficult and unnatural. 

One may wonder whether this result might change if more generic intersecting configurations are considered instead of simply taking parallel D-branes. In that case, the spatial directions are divided into NN, DN, ND, or DD groups. As shown in \cite{abel}, the single string density of states does not depend on ND and DN moduli, and the $f$-function in \eq{f} becomes
\be
f=n^2\fr{V_{NN}}{V_{DD}}.
\ee
Therefore, the total density of states also become independent of ND and  DN moduli, and pressures along these directions exactly vanish. Furthermore, the pressures along NN and DD directions are still ignorable compared to energy density, which shows that even in a generic situation involving intersecting branes one finds an effective pressure-less phase at string scale radii. Thus, the cosmic evolution and interaction rates do not change in the small radius regime and the assumption of thermal equilibrium is still questionable even intersecting D-branes exist.

Another possible point of concern is the effect of closed strings on the evolution and the interaction rates. As shown in \cite{b2}, the Boltzmann equations imply that in equilibrium the average length and thus the average energy of closed strings is suppressed by the number of open strings $\cn$, i.e. $L_c=L_o/\cn$ and $E_{c}=E_o/\cn$ (see eq. (17) in \cite{b2}). From \eq{cnc}, we have $\cn\gg1$ and thus the contributions of the closed strings on the cosmic evolution and interaction rates can safely be neglected due to the smallness of their energy and the length. 

Although we have discovered that the background \eq{sol1}-\eq{sol2}  is problematic, it is interesting to examine some further aspects of the solution. For instance, although the number of D-branes is taken to be small, one may want to inspect the issue of brane anti-brane annihilation. To be safe, one can simply require the Hubble expansion speed to be much larger than the peculiar brane speed, which can be assumed to be order one in string units.\footnote{Although strings are very energetic, most of their energies are stored in their masses and their motion can be viewed to be ``non-relativistic''. D-branes embedded in such a bath of strings are not expected to move fast or stay motionless.} Since the typical separation between D-branes is initially $n^{-1/d_D}$,  one thus demands 
\be\label{bs}
1\leq H\,n^{-1/d_D}.
\ee
Using \eq{hb}, this implies
\be\label{co2}
\fr{n^{2/d_D}}{E}\leq e^{\f_0},
\ee
which can only be satisfied marginally along  with  the Jeans instability condition \eq{co2}.  

As the Dirichlet directions expand, the expression for the entropy \eq{ss} looses its validity around $R_D\sim 3$ and one enters into a large radius regime. In the beginning of this regime we have\footnote{Actually one expects $V_N$ to be slightly larger than $V_D$, since \eq{efe} shows that the positive pressure along Neumann direction helps the expansion while the negative pressure along Dirichlet directions prevents it. Note that $K\dot{B}$ and $K\dot{C}$ terms play the role of a "velocity" dependent friction for $K>0$.} $V_N\simeq V_D$  and thus $f\geq1$. From \eq{lo} the entropy is now given by 
\be\label{el}
S\simeq \b_h E+a_0\,f\,V_D+2\sqrt{f(E-a_1\,f\,V_D)},
\ee
where the condition \eq{con2} is assumed to be satisfied (note that \eq{con} is automatically obeyed since $f\geq1$). Since $n={\cal O}(1)$, \eq{con2} is obeyed for $E\gg1$. From \eq{el}, the temperature and pressures can be found as  
\bea
&&\fr{1}{T}=\b_h+\sqrt{\fr{f}{E-a_1fV_D}},\nn\\
&&P_N=T\,a_0\,f\,V_D+T\sqrt{f(E-a_1\,f\,V_D)}-Ta_1fV_D\sqrt{\fr{f}{E-a_1fV_D}},\label{tpl}\\
&&P_D=-T\sqrt{f(E-a_1\,f\,V_D)}.\nn
\eea
As long as  \eq{con2} is obeyed, we have $E\gg f$ and $E\gg fV_D$, which shows that $T\simeq1/\b_h$ and $E\gg P_N,P_D$. Thus, the pressures are still negligible in this regime and the solution \eq{sol1}-\eq{sol2} is valid until \eq{con2} is violated. Actually, since the dilaton is increasing in \eq{sol2}, there is also a possibility that  the string coupling $e^\f$  can become ${\cal O}(1)$ and one enters in the strong coupling regime before even \eq{con2}  looses its validity. Whether this happens or not depends on the initial conditions. In any case, as long as the equations \eq{tpl} are trusted one ends up with an effectively pressureless evolution in this setup. 

At this point one may wonder why the arrow of time is chosen to yield an increasing dilaton and thus flow to a strong coupling regime. This choice actually dictated by the constraint equation in \eq{efe} as follows. According to the last equation, the sign of $K$ cannot alter during the evolution. Initially the time flow should be fixed to yield a positive $K$, since this choice produces "friction" terms in $B$, $C$ and $\f$ equations which avoids singularities. The initial conditions with $K<0 $ give run-away solutions usually developing a naked singularity in a finite proper time and thus these are not suitable for the description of the universe following big-bang. In our case, the choice $K>0$ gives an increasing dilaton. 

One may also consider  how the overall picture alters if the the number of D-branes $n$ is large and comparable to initial energy $n\sim E_0$. In that case, \eq{length} implies that $\bar{L}={\cal O}(1)$ and thus short strings dominate the system, which is due to  presence of a large number of D-branes chopping up the long strings. From \eq{tp}, one sees that the temperature is well below the Hagedorn temperature and pressures have the same order of magnitude with  energy.  From \eq{tp} defining an effective  equation of state parameter $\o$ as 
\be\label{lnd}
P_N=\o\,E,\hs{10}P_D=-\o\, E, \hs{10}
\o=\fr{1}{1+\b_h\sqrt{E/f}},
\ee
one finds that initially $\o\sim0.1$ if $n\sim E_0$. Since $\o$ is changing in time it is difficult to solve the field equations exactly or to utilize a simple approximation strategy. However, without even solving the system, one can observe  that there is an obstruction which arises due to D-brane anti-D-brane annihilation. Equation \eq{bs} implies that to avoid brane anti-brane collisions the Hubble expansion speed must be large
\be
1\ll E_0^{1/d_D}\sim n^{1/d_D} \leq H.
\ee
On the other hand the interaction rate in the short string phase can be estimated from \eq{sse} as
\be
\Gamma\sim  n\,\sqrt{E}\,e^{2\f}\sim e^{2\f_0}\, E_0^{3/2}\ll1,
\ee
where the last inequality follows from Jeans instability condition \eq{co1}. Therefore, $\Gamma\ll H$ and it seems difficult  to avoid Jeans instability, assume thermal equilibrium and  safely ignore D-brane annihilation process in this scenario. 

\section{Conclusions}\label{sec4}

In this paper, we consider  a toy cosmological model dominated by open strings attached to D-branes and determine the corresponding dilaton-gravity solution. We use basic thermodynamical properties of open strings on D-brane backgrounds to calculate the energy-momentum tensor, which turns out to be conserved under the assumption of adiabaticity.  Consistent coupling of this conserved energy-momentum tensor to dilaton-gravity background is determined using the contracted Bianchi identity. Contrary to closed strings, open strings attached to D-branes couple to dilaton and this alters the field equations in a significant way. Namely, the dynamical evolution of a direction is governed not only by the corresponding pressure as in the case of closed strings, but by the sum of pressure and energy density. Although the pressures are not small in string units, they can still be neglected compared to energy density and the early evolution is identical to a pressureless phase. Due to time depending dilaton, the metric functions differ from the usual matter dominated FRW cosmology. All directions and the dilaton tend to increase in the solution, which looses its validity until the energy density drops under a critical value or the strong coupling regime is reached. 

We check the self-consistency of the solution for a few issues. Firstly, avoiding Jeans instability imposes a constraint on the initial values of the dilaton and energy \eq{co1}. Secondly, to fulfill thermal equilibrium,  initial conditions must be finely tuned such that  the Hubble parameter becomes less than the interaction rate. However, even after this fine tuning, the Hubble parameter  increases to a value \eq{hb} larger than the interaction rate, provided the condition \eq{co1} is assumed to avoid Jeans instability. Therefore, as for closed strings, the assumption of thermal equilibrium is questionable and one requires  an understanding of non-equilibrium thermodynamics. On the other hand,  the existence of suitable initial conditions, even though they are finely tuned, shows that the cosmology of open string gases may differ substantially from that of closed strings. 

In the context of string/brane gas cosmology, a complete and plausible scenario is still absent. Therefore, the role that can be played by D-branes and open strings attached to them is not clear in the big picture. It is somehow discouraging to observe that open strings suffer from a crucial  puzzle encountered for closed strings. However, it looks like new features can arise that can modify the whole story. For instance, accepting the fine tuning in the initial conditions, the passage from thermal equilibrium to freezing out resembles the BV mechanism, so it would be interesting to dwell on this aspect by also including closed strings into the picture. Moreover, in such a scenario the Hubble parameter, which is initially vanishing,  increases in time for a finite duration giving a decreasing Hubble radius. Such a behavior is desirable for the structure formation mechanism proposed in \cite{nbv}, therefore it is of interest to analyze revisions which can possibly arise by the inclusion of D-branes.

\end{document}